\documentclass[aps,prl,reprint,twocolumn,showkeys,showpacs,preprintnumbers,amsmath,amssymb,floatfix,superscriptaddress]{revtex4-1}

\usepackage{graphicx}
\usepackage{dcolumn}
\usepackage{bm}
\usepackage{dsfont}

\begin{document}
\title{Electrical Detection of Coherent Nuclear Spin Oscillations in Phosphorus-Doped Silicon Using Pulsed ENDOR}

\author{Felix Hoehne}
\email[corresponding author, email: ]{hoehne@wsi.tum.de} \affiliation{Walter
Schottky Institut, Technische Universit\"{a}t M\"{u}nchen, Am Coulombwall 4,
85748 Garching, Germany}

\author{Lukas Dreher}
\email[corresponding author, email: ]{dreher@wsi.tum.de}
\affiliation{Walter Schottky Institut, Technische Universit\"{a}t
M\"{u}nchen, Am Coulombwall 4, 85748 Garching, Germany}

\author{Hans Huebl}
\affiliation{Walther-Mei\ss ner-Institut, Bayerische Akademie der Wissenschaften, Walther-Mei\ss ner-Str.\,8, 85748 Garching, Germany}

\author{Martin Stutzmann}
\affiliation{Walter Schottky Institut, Technische Universit\"{a}t
M\"{u}nchen, Am Coulombwall 4, 85748 Garching, Germany}

\author{Martin S.~Brandt}
\affiliation{Walter Schottky Institut, Technische Universit\"{a}t
M\"{u}nchen, Am Coulombwall 4, 85748 Garching, Germany}
\begin{abstract}
We demonstrate the electrical detection of pulsed X-band Electron Nuclear Double Resonance (ENDOR) in phosphorus-doped silicon at 5\,K. A pulse sequence analogous to Davies ENDOR in conventional electron spin resonance is used to measure the nuclear spin transition frequencies of the $^{31}$P nuclear spins, where the $^{31}$P electron spins are detected electrically via spin-dependent transitions through Si/SiO$_2$ interface states, thus not relying on a polarization of the electron spin system. In addition, the electrical detection of coherent nuclear spin oscillations is shown, demonstrating the feasibility to electrically read out the spin states of possible nuclear spin qubits.
\end{abstract}


\maketitle


Control and readout of electron spin states in solids are well established down to the single spin level \cite{hanson_spins_2007,Morello10}. In contrast, the readout of nuclear spin states has mostly been limited to optical techniques \cite{jelezko_nuc_osc_observation_2004,Childress06NVQC}. However, for nanostructures not exhibiting luminescence, an electrical readout scheme is required. Electrical detection of nuclear magnetic resonance has been performed on e.g.  two-dimensional electron gases, identifying the origin of the Overhauser field \cite{dobers_electrical_1988}, and the electrical readout of nuclear spin states has been achieved for this system \cite{yusa_controlled_2005}. In phosphorus-doped silicon,  McCamey {\it et al.} \cite{mccamey_electronic_2010} have recently demonstrated the electrical readout of nuclear spins at 8.6\,T. While both these studies employ highly polarized spin systems for the readout, we make use of a spin-dependent recombination process via Si/SiO${_2}$ interface states \cite{Hoehne09KSM}; this approach does not rely on a polarization of the electron spin system and thus works under experimental conditions where the thermal energy is much larger than the electron Zeeman splitting \cite{Kaplan78Spindep}. We here demonstrate the electrical detection of coherent nuclear spin oscillations in phosphorus-doped silicon using pulsed Electrically Detected Electron Nuclear Double Resonance (EDENDOR) at 0.3~T using X-band frequencies (10 GHz). Outside the framework of quantum computation, pulsed EDENDOR could become a valuable spectroscopic tool for the characterization of point defects in semiconductors combining the high sensitivity of continuous-wave EDENDOR \cite{stich_electrical_1996} with the reduction of the dynamic complexity of the coupled electron and nuclear spin systems via pulsed spectroscopy \cite{gemperle_pulsed_1991}.

\begin{figure*}[!htb]
\begin{centering}
\includegraphics{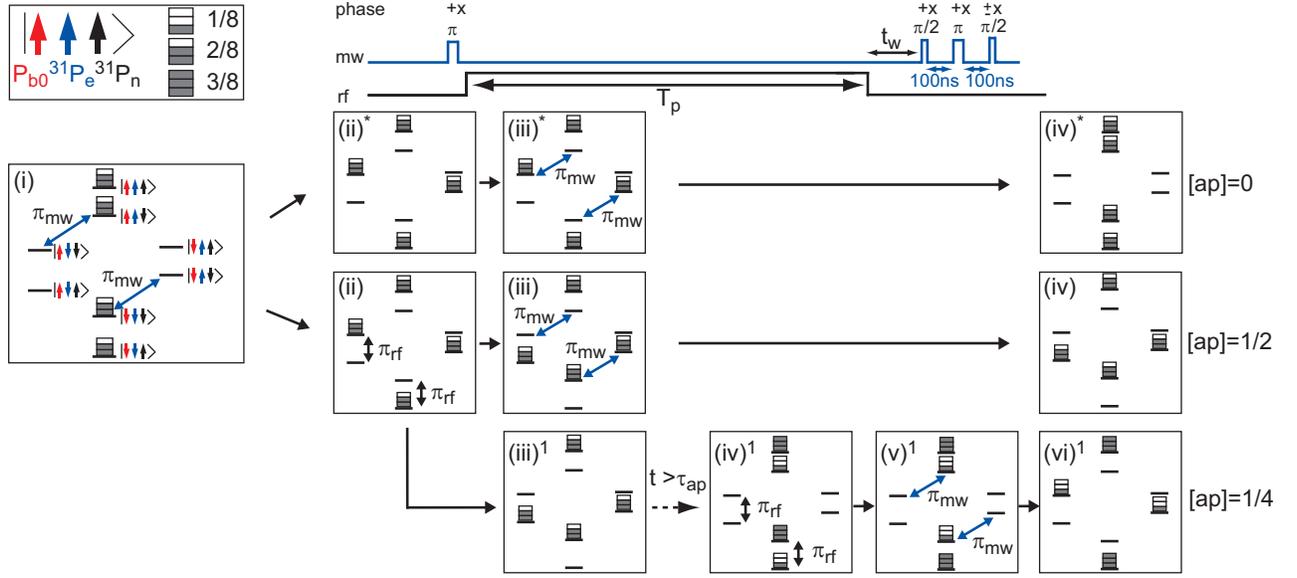}
\par\end{centering}
\caption{\label{fig:Didaktik} The pulse sequence of the electrically detected Davies ENDOR consists of a preparation microwave-$\pi$-pulse, an rf pulse of length $T_\mathrm{p}$, and a mw detection $\pi$-pulse after a waiting time $t_\mathrm{w}$. This pulse is replaced by a mw detection echo-sequence, where the final $\pi$/2-pulse is phase shifted by 180 degrees with respect to the other pulses to form an effective $\pi$-pulse. The evolution of the spin state populations for an ideal EDENDOR is depicted in the panels in the upper part of the figure. The donor electron spin ($^{31}$P$_\mathrm{e}$), its nuclear spin ($^{31}$P$_\mathrm{n}$), and the electron spin of the Si/SiO$_2$ interface state (P$_{\rm{b0}}$) are symbolized by the three arrows in the bracket. The populations of the spin states are indicated by gray boxes. The fraction of antiparallel spin pairs at the end of the pulse sequence is given by $[ap]$. Panels (iii)$^1$-(vi)$^1$ illustrate the evolution of the populations when recombination of antiparallel spin pairs during the pulse sequence is taken into account.}
\end{figure*}

The basic principle of pulsed EDENDOR is depicted in Fig.~\ref{fig:Didaktik}. For the $^{31}$P donor in silicon to be investigated, its electron spin $^{31}\mathrm P_\mathrm{e}$ with $S=1/2$, its nuclear spin $^{31}\mathrm P_\mathrm{n}$ with $I=1/2$, and their hyperfine coupling give rise to a four-level system. For the electrical readout we use a spin-to-charge conversion mechanism based on a spin-dependent recombination involving the $^{31}$P donor electron and the P$_\mathrm{b0}$ center ($S=1/2$) at the Si/SiO$_2$ interface \cite{Hoehne09KSM}. Accounting for the two orientations of the P$_\mathrm{b0}$ spin, we sketch the eight different states for the three involved spins in panel (i), indicating the occupation of the different states by the gray bars. Due to the Pauli principle, spin pairs with antiparallel spins recombine while the pairs with parallel orientation are long-lived. Therefore, in the steady state only levels associated with parallel orientations of $^{31}$P$_\mathrm{e}$-P$_\mathrm{b0}$ pairs are occupied \cite{Kaplan78Spindep} with a probability of 1/4 each. We neglect the polarization of the spin system at first.

The presented pulsed EDENDOR experiments are based on the Davies pulse sequence for conventional pulsed ENDOR \cite{davies_new_1974}. At first, we describe the EDENDOR experiments assuming that no recombination occurs on the timescale of the pulse sequence. The preparation microwave (mw) $\pi$-pulse inverts the populations of the levels associated with one of the two $^{31}$P$_\mathrm{e}$ hyperfine transitions (i). Subsequently, a radiofrequency (rf) $\pi$-pulse inverts the populations on one of the $^{31}$P$_\mathrm{n}$ hyperfine transitions (ii). Since the signal observed in pulsed electrically detected magnetic resonance is proportional to the fraction $[ap]$ of $^{31}$P$_\mathrm{e}$-P$_\mathrm{b0}$ pairs with antiparallel spins at the end of the pulse sequence \cite{Boehme03EDMR}, a detection mw-$\pi$-pulse is applied to electrically readout the nuclear spin state (iii). At the end of this pulse sequence, we expect an antiparallel spin fraction of $[ap]=1/2$ when an rf $\pi$-pulse has been applied (iv) whereas $[ap]=0$ without an rf pulse or with an rf pulse far off resonance (iv)*. In contrast to conventional ENDOR, where the ENDOR intensity is limited by the polarization of the electron spin ensemble, the EDENDOR intensity is given by $[ap]$, which can be significantly larger. 



For the measurements of EDENDOR, we used a sample consisting of a $22$\,nm-thick Si layer with a nominal P concentration of $9\times10^{16}$\,cm$^{-3}$ grown by chemical vapor deposition on a $2.5\,\mu$m thick, nominally undoped Si buffer on a silicon-on-insulator substrate. The epilayer is covered with a native oxide. Interdigit Cr/Au contacts with a periodicity of $20\,\mu$m defining an active area of $2\times2.25$\,mm$^{2}$ were evaporated for photoconductivity measurements. The sample was mounted with the silicon [110] axis parallel to the static magnetic field, illuminated with above-bandgap light, and biased with 100\,mV resulting in a current of $\approx$60$\mu$A. 
All experiments were performed at $\approx$5\,K in a BRUKER dielectric microwave resonator for pulsed X-band ENDOR. The microwave-pulse power was adjusted such that the $\pi$-pulse length was 30\,ns, corresponding to a microwave $B_1$-field of 0.6\,mT. The current transients after the pulse sequence were recorded by measuring the voltage drop over a 1.6\,k$\Omega$ resistor placed in series with the sample. The resulting voltage transients were filtered by a cascade of low-pass filters ($f_{\mathrm{3dB}}\approx4$\,MHz), amplified, and recorded with a digitizer card. The pulse sequence used is shown on the top of Fig.~\ref{fig:Didaktik}. To remove the background resulting from non-resonant photocurrent transients\,\cite{Stegner06}, we replace the final detection mw-$\pi$-pulse by a detection echo-sequence \cite{jelezko_nuc_osc_observation_2004,Huebl08Echo}; by applying a two-step phase cycle \cite{Schweiger01} to the last microwave $\pi/2$-pulse (indicated in Fig.~\ref{fig:Didaktik} by $\pm x$), with the phase changed by 180 degrees for every shot, and subtracting consecutive traces, only resonant changes in the photocurrent are detected. To obtain a sufficient signal-to-noise ratio, the experiment is repeated with a shot repetition time of 800 $\mu$s allowing the electron spin system to relax to its steady-state\,\cite{Huebl08Echo, Paik2010T1T2}. The recorded current transients are box-car integrated, resulting in a charge $\Delta Q$ that is proportional to the antiparallel spin pair fraction $[ap]$ at the end of the microwave pulse sequence\,\cite{Boehme03EDMR}.  
               
\begin{figure}[h]
\begin{centering}
\includegraphics{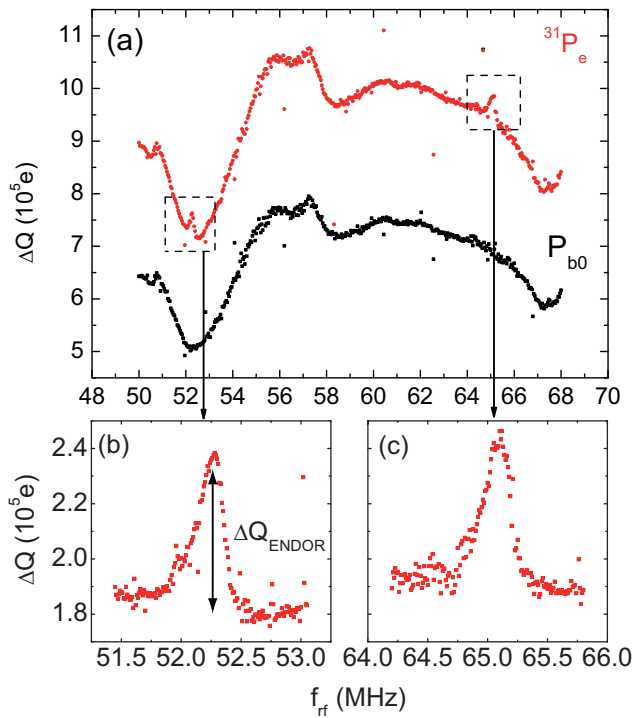}
\par\end{centering}
\caption{\label{fig:Spektrum}
(a) Integrated current transient $\Delta Q$ after the Davies EDENDOR pulse sequence with $T_{\rm{p}}$=10\,$\mu$s as a function of the rf frequency. When the mw pulses are in resonance with the $^{31}$P high-field hyperfine line (upper trace) two peaks can be observed at $f_{\rm{rf}}$=52.25 MHz and $f_{\rm{rf}}$=65.08 MHz. These peaks do not appear in the lower trace where the mw pulses are in resonance with the P$_{\rm{b0}}$ spins. More detailed frequency scans of the two peaks are shown in panels (b) and (c), where the non-resonant background has been subtracted. $\Delta Q_\mathrm{ENDOR}$ denotes the peak amplitude.}
\end{figure}
%
In a first experiment, we demonstrate that pulsed EDENDOR can be used as a spectroscopic method. To this end, we keep the length of the rf pulse fixed at $T_{\mathrm{p}}$=10\,$\mu$s while sweeping the rf frequency. For the remainder of this work, the detection echo-sequence with equal evolution times of 100~ns is applied after waiting for $t_{\mathrm{w}}$=2\,$\mu$s, c.f. Fig.~\ref{fig:Didaktik}. Figure~\ref{fig:Spektrum} (a) shows the integrated current transient $\Delta Q$ as a function of the rf frequency $f_\mathrm{rf}$ for two different magnetic fields, chosen such that the microwave pulses are resonant with the high-field hyperfine-split $^{31}$P$_\mathrm{e}$ resonance (upper trace) and with the  P$_{\rm{b0}}$ centers at $g$=2.0042 (lower trace) \cite{Stesmans98}. Since for non-resonant rf pulses the antiparallel spin content is the same for microwave pulses resonant with the $^{31}$P$_\mathrm{e}$ and with the P$_{\rm{b0}}$ spins, comparison of the two traces allows to identify the $^{31}$P$_\mathrm{n}$ transitions at frequencies of $f_1$=52.25$\pm$0.02\,MHz and $f_2$=65.08$\pm$0.02\,MHz. More detailed frequency scans of the two peaks are shown in panels (b) and (c) where the P$_{\rm{b0}}$ trace was subtracted from the $^{31}$P$_\mathrm{e}$ trace to remove the non-resonant background signal. The measured nuclear transition frequencies are in good agreement with the expected frequencies of 52.34\,MHz and 65.19\,MHz for the hyperfine interaction of 117.53\,MHz between $^{31}$P$_\mathrm{e}$ and $^{31}$P$_\mathrm{n}$, and the $^{31}$P$_\mathrm{n}$ nuclear Larmor frequency of 6.076\,MHz \cite{Feher59I}. The slightly smaller value of the measured frequencies could be attributed to small deviations of the hyperfine interaction at the surface from the bulk value due to local strain in the thin-film silicon sample\,\cite{Huebl2006Strain}. 
To exclude spurious effects of the intense rf pulse, we repeated the same measurement without the preparation mw-$\pi$-pulse. As expected, no peaks at the nuclear transition frequencies were observed (data not shown). 
\begin{figure}[h]
\begin{centering}
\includegraphics{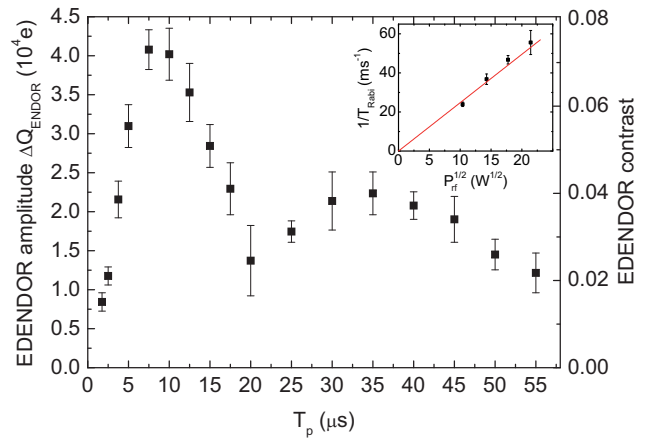}
\par\end{centering}
\caption{\label{fig:NuclearRabis}
Amplitude of the EDENDOR peak $\Delta Q_\mathrm{ENDOR}$ at 52.25\,MHz as a function of the rf pulse length $T_{\mathrm{p}}$. The oscillation shows the coherent driving of nuclear spin motion which is damped on the timescale of the recombination of the $^{31}$P$_\mathrm{e}$-P$_{\rm{b0}}$ spin pairs. The Rabi frequency scales linearly with the square root of the rf power, as shown in the inset.}
\end{figure}
%

To demonstrate the electrical readout of coherent nuclear spin oscillations, we measured the amplitude of the EDENDOR signal as a function of the rf pulse length $T_\mathrm{p}$ in a second experiment. To this end, we recorded rf frequency sweeps as those in Fig.\,\ref{fig:Spektrum} (b) for different $T_\mathrm{p}$ and fitted the peaks with Lorentzians. In Fig.\,\ref{fig:NuclearRabis}, we show their amplitudes $\Delta Q_\mathrm{ENDOR}$ as a function of $T_\mathrm{p}$, revealing a damped oscillation. We attribute this oscillation to the coherent driving of the $^{31}$P nuclear spins. To corroborate this interpretation, we measured the oscillation period $T_{\mathrm{Rabi}}$ for different rf power levels $P_{\mathrm{rf}}$ resulting in a linear increase of the Rabi frequency $1/T_{\mathrm{Rabi}}$ with the rf $B_2$-field ($B_2\sim \sqrt{P_{\mathrm{rf}}}$) as shown in the inset of Fig.\,\ref{fig:NuclearRabis}. To compare the measured EDENDOR amplitude $\Delta Q_\mathrm{ENDOR}$ with the theoretically expected maximum value $[ap]=1/2$ from the considerations depicted in Fig.~\ref{fig:Didaktik}, we take the amplitude $\Delta Q_\mathrm{echo}$ of a reference detection echo defined above (which corresponds to an effective $\pi$-pulse) without preceding mw and rf pulses under the same experimental conditions as a measure for the maximum experimentally achievable value of $[ap]$. We define the EDENDOR contrast as $\Delta Q_\mathrm{ENDOR}/\Delta Q_\mathrm{echo}$, resulting in a contrast of 0.07 at $T_\mathrm{p}$=10\,$\mu$s.

The evolution of the spin system during the Davies EDENDOR sequence deviates in several aspects from the ideal situation discussed above. These are primarily the recombination that occurs during the rf pulse and the polarization and relaxation of the electron and nuclear spins. A full simulation of the experimentally observed data will require incorporation of these dynamic effects e.g. in a system of combined rate equations \cite{Boehme03EDMR}. To gain a first physical picture, we here separately discuss the effects of the recombination time $\tau_{\mathrm{ap}}$ characteristic for antiparallel electron spin pairs, of the recombination time $\tau_{\mathrm{p}}$ characteristic for the parallel spin pairs, and of the polarization of the nuclear spin system. To independently quantify the recombination time $\tau_\mathrm{ap}$, we performed an inversion-recovery type of experiment\,\cite{Paik2010T1T2}. The observed decay can be described with a stretched exponential with a time constant of $\approx 7 \mu$s, which we identify with $\tau_{\mathrm{ap}}$, and an exponent of $\approx 0.5$. (The recombination time $\tau_{\mathrm{p}}$ is not readily accessible; we assume that it is much longer than $\tau_{\mathrm{ap}}$ due to the weak spin-orbit coupling in Si and that is shorter than the shot repetition time used.) Thus, after an rf pulse with a length of $T_\mathrm{p}$=10\,$\mu$s, we expect an EDENDOR amplitude of $0.3\cdot \Delta Q_\mathrm{echo}$. In addition, the finite excitation bandwidth of the rf pulse only excites about a fraction of 0.6 of the nuclear spin resonance line with a FHWM of 0.12\,MHz. This reduces the expected EDENDOR contrast to $\approx$0.18, which agrees within better than a factor of 3 with the observed EDENDOR contrast despite the crude approximations involved.

As Fig.~\ref{fig:NuclearRabis} demonstrates\label{page4}, even for times $T_\mathrm{p}$ much longer than $\tau_\mathrm{ap}$ we observe an EDENDOR oscillation amplitude. This can be qualitatively explained in terms of the evolution of the spin system depicted in Fig.~\ref{fig:Didaktik}. To estimate the effect of longer rf pulses $T_\mathrm{p}$, we hypothetically divide $T_\mathrm{p}$ into an ideal $\pi$ pulse without recombination during the pulse, a time period $t$, during which the antiparallel spin pairs recombine, and an additional ideal rf pulse. Starting from the spin system after the first ideal $\pi$ rf pulse (iii)$^1$, we consider the situation after a time $t$ ($\tau_\mathrm{p}\!>t\!>\!\tau_\mathrm{ap}$) when all antiparallel spin pairs have recombined leaving unoccupied donor states. These are refilled with electrons that form new parallel spin pair states with equal probability for each of the four parallel spin-pair states (iv)$^1$, since the formation of new spin pairs is not spin-dependent \cite{Kaplan78Spindep}. The additional ideal rf pulse induces nuclear spin oscillations that can be measured by a detection echo because of the remaining population difference. For example, after an additional rf $\pi$-pulse we expect $[ap]=1/4$ as shown in (iv)$^1$-(vi)$^1$, whereas an additional $2\pi$ rf pulse would result in $[ap]=1/2$. Therefore, the expected amplitude of the oscillation is reduced by a factor of 2 compared to the oscillations expected according to (ii)-(iv). Motivated by the above considerations, we tentatively attribute the experimentally observed persistence of the oscillation and its amplitude at pulse lengths $T_\mathrm{p}\!\gg\! \tau_\mathrm{ap}$ to the substantial difference in the involved recombination times.  These oscillations are expected to decay on the timescale of $\tau_\mathrm{p}$, which would be accessible by EDENDOR according to this line of argument.

In analogy to conventional Davies ENDOR \cite{davies_new_1974}, the ideal EDENDOR pulse sequence transfers the electron spin polarization to the nuclei. This transfer of polarization allows to readout the nuclear spin state after waiting times $t_\mathrm{w}$ much larger than the recombination times, limited only by $T_{\mathrm{1n}}$. Indeed, we observe an EDENDOR signal for $t_\mathrm{w}>4~\mathrm{ms}$ with an EDENDOR contrast of 0.01 using a shot-repetition time of 15~ms. This would allow to measure $T_\mathrm{1n}$ by recording the EDENDOR contrast as a function of $t_\mathrm{w}$.

In conclusion, we have demonstrated the electrical detection of coherent nuclear spin oscillations with pulsed EDENDOR employing Si:P as a model system. We use a spin-dependent recombination process via Si/SiO${_2}$ interface states enabling the nuclear spin readout without making use of a polarization of the electron spin system. During the readout process, the electron is removed from the donor, which drastically changes the Hamiltonian describing the spin system. For applications of this approach as quantum memory \cite{morton_solid-state_2008,mccamey_electronic_2010}, such effects might be suitably suppressed by decoupling the $^{31}$P$_\mathrm{e}$-P$_\mathrm{b0}$ readout pair using local gate electrodes, where an application of an electric potential keeps the donor in its paramagnetic state for a long time. On the other hand, it would be interesting to study the effects of removing and repopulating the electron on the nuclear spin state in detail, e.g. with respect to the relaxation and decoherence of the nuclear spin system. Furthermore, pulsed EDEDNOR gives access to the recombination time of parallel spin pairs that has hitherto eluded unambiguous experimental determination as well as the nuclear relaxation time \cite{mccamey_electronic_2010}. The high sensitivity of the electrical detection of magnetic resonance and the widespread availability of X-Band spectrometers makes the presented approach to pulsed EDENDOR an attractive spectroscopic method for the characterization of defects in semiconductor nanostructures. Hereby, hyperfine and superhyperfine interactions can be investigated, which allow to obtain quantitative information about the electron wavefunction and the specific environment around the defect. Since it does not rely on polarization of the spin system, this approach could also be particularly attractive for room temperature EDENDOR spectroscopy, e.g. of endohedral fullerenes such as N@C60 \cite{harneit_room_2007}.


\begin{acknowledgments}
The work was supported by DFG
(Grant No. SFB 631, C3).
\end{acknowledgments}

\end{document}